\documentclass[aps,prb,reprint,longbibliography,superscriptaddress]{revtex4-1}

\usepackage{amsmath,amsfonts,amssymb}
\usepackage{graphicx}
\usepackage{color}
\usepackage{bm}
\usepackage[dvipsnames]{xcolor}
\usepackage{braket}
\usepackage{epstopdf}
\usepackage{enumitem}
\usepackage[colorlinks]{hyperref}

\let\ifr\i

\renewcommand{\i}{{\rm i}}

\renewcommand{\Re}{\mathop{\rm Re}}

\begin{document}

\title{Spin Inertia of Resident and Photoexcited Carriers in Singly-Charged Quantum Dots}

\author{E.~A. Zhukov}
\affiliation{Experimentelle Physik 2, Technische Universit\"at Dortmund, 44221 Dortmund, Germany}

\author{E. Kirstein}
\affiliation{Experimentelle Physik 2, Technische Universit\"at Dortmund, 44221 Dortmund, Germany}

\author{D.~S. Smirnov}
\affiliation{Ioffe Institute, Russian Academy of Sciences, 194021 St.\,Petersburg, Russia}

\author{D.~R. Yakovlev}
\affiliation{Experimentelle Physik 2, Technische Universit\"at Dortmund, 44221 Dortmund, Germany}
\affiliation{Ioffe Institute, Russian Academy of Sciences, 194021 St.\,Petersburg, Russia}

\author{\\M.~M. Glazov}
\affiliation{Ioffe Institute, Russian Academy of Sciences, 194021 St.\,Petersburg, Russia}

\author{D. Reuter}
\affiliation{Universit\"at Paderborn, Department Physik, 33098 Paderborn, Germany}

\author{A.~D. Wieck}
\affiliation{Angewandte Festk\"orperphysik, Ruhr-Universit\"at Bochum, 44780 Bochum, Germany}

\author{M. Bayer}
\affiliation{Experimentelle Physik 2, Technische Universit\"at Dortmund, 44221 Dortmund, Germany}
\affiliation{Ioffe Institute, Russian Academy of Sciences, 194021 St.\,Petersburg, Russia}

\author{A. Greilich}
\affiliation{Experimentelle Physik 2, Technische Universit\"at Dortmund, 44221 Dortmund, Germany}

\begin{abstract}
The spin dynamics in a broad range of systems can be studied using circularly polarized optical excitation with alternating helicity. The dependence of spin polarization on the frequency of helicity alternation, known as the
spin inertia effect, is used here to study the spin dynamics in singly-charged (In,Ga)As/GaAs quantum dots (QDs) providing insight into spin generation and accumulation processes. We demonstrate that the dependence of spin
polarization in $n$- and $p$-type QDs on the external magnetic field has a characteristic V- and M-like shape, respectively. This difference is related to different microscopic mechanisms of resident carriers spin orientation. It allows us to determine the parameters of the spin dynamics both for the ground and excited states of singly-charged QDs.
\end{abstract}

\maketitle

Long carrier spin coherence and spin relaxation times are the main prerequisites for a system to be suited for quantum information technologies~\cite{Loss1998,LaddNature10}. The main route to extend
these times is to isolate the studied system from its environment. At the same time, such isolation reduces the possibility for a fast state readout and manipulation. Carrier spins confined in low-dimensional semiconductor
structures, in particular, singly-charged (In,Ga)As quantum dots (QDs), offer a unique balance of accessibility and robustness~\cite{henneberger2016semiconductor}. The direct optical band gap with a giant optical dipole moment of a semiconductor QD~\cite{dyakonov_book} allows exploiting the optical excitation, the exciton, as an auxiliary state for ultra-fast conversion between the optical coherence of a laser pulse (picosecond duration) and the long-living spin coherence of an isolated resident carrier (microsecond coherence). This opportunity has triggered considerable activity in optical operations with QDs, including spin initialization~\cite{Atature06}, nondestructive readout~\cite{Kim2008}, and fast spin manipulation~\cite{RamsayReview}.

The way the optical coherence is transferred to the spin-coherence in a QD involves the process of optical orientation and excitation of a trion, the exciton which is bound to the resident carrier. After
excitation, the trion recombines stochastically and leaves behind a polarized single carrier spin in the ground state, which is measured. Generally, the spin relaxation dynamics within a trion determines the final
spin polarization. The standard optical methods to access the relaxation dynamics of resident spins are the Hanle effect~\cite{Hanle} and time-resolved pump-probe techniques~\cite{AwschalomSpintronics}. However, these techniques provide limited access to the spin generation process and dynamics of the photoexcited states~\cite{dyakonov_book}.

In this paper, we access the parameters of the spin dynamics in the trion employing the recently developed \textit{Spin Inertia} (SI) technique~\cite{Heist2015}. In SI, the spin dynamics in the system is related to the helicity modulation of the excitation polarization~\cite{Ignatiev05,Akimov2006,FrasPRB11}. The maximal value of the spin polarization, created by the circularly polarized pump pulses, is then traced as a function of the helicity modulation frequency $f_m$. The spin polarization stays constant for $f_m \lesssim T_s^{-1}$, where $T_s$ is the spin relaxation time of a carrier, and becomes reduced if the modulation frequency is larger than $T_s^{-1}$. The cut-off frequency defines the spin relaxation time~\cite{Heist2015,TSI}.

In the presented experiment we study the spin relaxation in a model system, namely an ensemble of singly-charged (In,Ga)As/GaAs QDs, doped either with resident electrons or holes. The maximal spin polarization is determined by the interplay of the carrier spin generation and relaxation dynamics. Importantly, applying a longitudinal magnetic field, a substantial difference in the magnetic field behavior is observed for resident electron and hole spins. The origin of the observed difference and its relation to the SI represent the main topic of the paper. Using the developed theory, we are able to exploit the full potential of the experiment. We describe the shape of the magnetic-field-dependent traces and extract information on the photoexcited carrier spin dynamics, which can be accessed from the signal accumulated long after the radiative trion decay. Furthermore, we determine spin relaxation times, longitudinal $g$ factor values and hyperfine interaction strengths in the ground and excited states of the system.

We study $n$- and $p$-doped ensembles of singly-charged self-assembled QDs. Both samples contain 20 layers of molecular beam epitaxy grown (In,Ga)As QDs separated by 60\,nm GaAs barriers. The average QD density is about $10^{10}$\,cm$^{-2}$ per layer. The $p$-doped sample has a background level of $p$-type doping due to residual carbon impurities. The $n$-doped sample was obtained by incorporating $\delta$-sheets of Si 20\,nm below each QD layer. The samples are mounted in the variable temperature insert of a magneto-optical bath cryostat ($T=1.5 - 300$\,K), and are excited by the laser close to the maximum of the photoluminescence at 1.412\,eV (878\,nm) for the $n$-type QDs, and at 1.392\,eV (891\,nm) for the $p$-type QDs~\cite{GlasenappPRB16}. The laser spot diameters on the sample are 300\,$\mu$m. As the samples are different in their doping types and concentrations, different excitation conditions are required for strongest signal level. The magnetic field $\bm B$ is applied in Faraday geometry, along the optical $z$-axis. Details on the sample characterization are given in Refs.~[\onlinecite{Varwig2012,chapter6}].

The spin polarization is created by circularly polarized pump pulses of 1.5\,ps duration emitted by a mode-locked Ti:Sapphire laser operating at a repetition period $T_R = 13.2$\,ns. The pump helicity is modulated at frequency $f_m$ between $\sigma^+$ and $\sigma^-$ polarization by an electro-optical modulator (EOM). Linearly polarized probe pulses measure the induced spin polarization along the optical axis via the ellipticity signal in transmission geometry analyzed by a quarter wave plate and a wollaston prism, using a balanced diode bridge~\cite{Heist2015}. This signal is called Faraday Ellipticity (FE).

In the presented experiments, we fix the pump-probe delay at a negative value ($\tau_{pp}=-50$\,ps) and measure the FE dependence on the longitudinal magnetic field, applied along the excitation direction and orthogonal to the
sample surface. Using the signal at negative time delay greatly simplifies the interpretation of the results, as it can only arise from the spin polarization of the resident long-living carriers. In the studied QDs the trion recombination time of 400\,ps is much shorter than $T_R$~\cite{GreilichPRB73}.

\begin{figure}[t]
\includegraphics[width=\columnwidth]{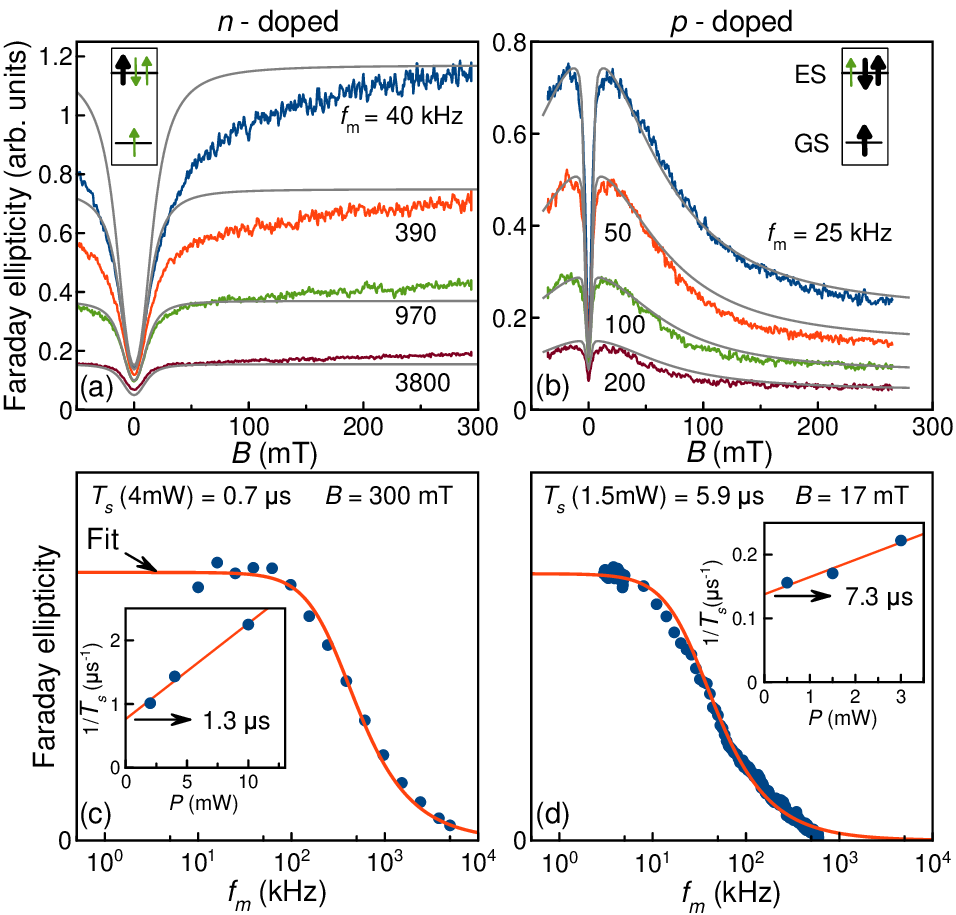}
\caption{Spin polarization of carriers as a function of magnetic field at different modulation frequencies of pump helicity: (a) $n$-doped sample, laser pump/probe power is $4/0.8$\,mW, $T=6$\,K; (b)
$p$-doped sample, pump/probe power is $1.5/0.5$\,mW, $T=1.7$\,K. Gray curves are calculations after Eq.~\eqref{eq:L} with parameters given in Tab.~\ref{tab:params}~\cite{params}. Insets show examples of possible ground
(GS) and excited state (ES) spin configurations: thin green and thick black arrows represent the electron and hole spins, respectively. Panels (c) and (d) show the frequency dependencies of carrier spin polarizations at
fixed magnetic field for the $n$-doped and $p$-doped samples, respectively. Red line is a fit by Eq.~\eqref{fit}. Insets are linear extrapolations of the power dependent spin lifetime $T_s$ down to zero pump power giving
the spin relaxation times in equilibrium, $T_s^{(0)}$.}
\label{fig:1}
\end{figure}

Figures~\ref{fig:1}(a) and~\ref{fig:1}(b) demonstrate the measured FE as a function of magnetic field for different modulation frequencies for $n$- and $p$-doped samples. We call this dependence a polarization recovery curve (PRC). For both samples the FE increases with increase of magnetic field around $B=0$. In the case of strong carrier localization, the dominant spin relaxation mechanism is the hyperfine interaction with host lattice nuclear spins. Application of the external magnetic field leads to an effective decoupling of the resident carrier spin from the nuclear spin bath and stabilizes the optically oriented spins along the $z$-axis~\cite{Heist2015,PetrovBoxModel09}, so FE increases. At higher magnetic fields the behavior is different: The electron spin polarization saturates (in the range around 300\,mT), while the hole spin polarization decreases after the initial increase and then saturates within a similar range of fields. The observed difference in the field dependencies of PRC is related to the type of carriers (electrons vs. holes) and the spin relaxation dynamics in the trion states, which are schematically shown in the insets in Figs.~\ref{fig:1}(a) and~\ref{fig:1}(b).

The SI method gives access to the relaxation dynamics of the resident carriers, as shown by the exemplary data sets in Figs.~\ref{fig:1}(c) and~\ref{fig:1}(d). Figure~\ref{fig:1}(c), shows the $f_m$ dependence of the FE at fixed magnetic field ($B=300$\,mT), extracted from the data shown in Fig.~\ref{fig:1}(a). Using the fit with the form~\cite{Heist2015}:
\begin{equation}
\label{fit}
\frac{FE(f_m)}{FE(0)} = \frac{1}{\sqrt{1+(2 \pi f_m T_s)^2}},
\end{equation}
we determine the characteristic spin lifetime $T_s=0.7$\,$\mu$s of the resident electrons at the pump power of 4\,mW~\cite{Heist2015}. In this case the spin polarization is far below the saturation level. The case of strong pumping is discussed in Ref.~[\onlinecite{TSI}]. The value of $T_s$ depends on the pump power due to the influence of the photoexcitation~\cite{Heist2015,LiPRL12}, hence, to determine the longitudinal spin
relaxation time in equilibrium, $T_s^{(0)}$, we apply different pump powers $P$ and extrapolate $T_s$ down to zero power using a linear fit to the inverse of the data, see inset in Fig.~\ref{fig:1}(c). It gives $T_s^{(0)}=1.3$\,$\mu$s for $B=300$\,mT in $n$-doped QDs. Figure~\ref{fig:1}(d) shows the corresponding frequency dependence for the $p$-doped sample around the maximum of the FE signal (at $B=17$\,mT) and $P=1.5$\,mW. The procedure described above yields $T_s=5.9\,\mu$s and $T_s^{(0)}=7.3$\,$\mu$s. This demonstrates the ability of SI to access very long spin relaxation times, exceeding $T_R$ by almost three orders of magnitude. Spin relaxation times in the microsecond range are in good agreement with other measurements~\cite{FrasPRB11,LiPRL12,GlasenappPRB16}, while the difference from the results of Ref.~[\onlinecite{Heiss2007}] can be explained by the difference in the experimental protocol~\cite{NoteX}.

\begin{figure}[t]
\includegraphics[width=\columnwidth]{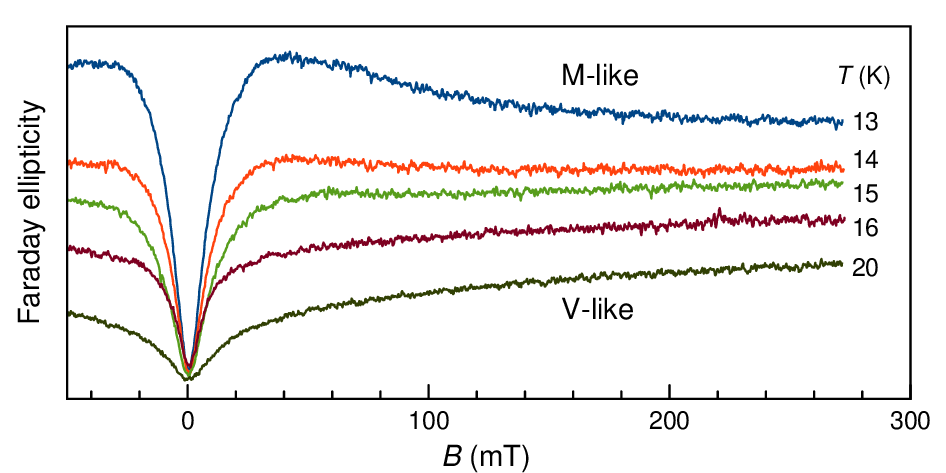}
\caption{Temperature dependence of PRC for the $p$-doped sample. Laser pump/probe power is $2/0.5$\,mW, $f_m=2$\,kHz.}
\label{fig:2}
\end{figure}

Importantly, the PRCs have different shapes for $n$- and $p$-doped samples. This situation changes with increasing temperature. Figure~\ref{fig:2} demonstrates the evolution of the PRC for the $p$-doped sample at low
modulation frequency $f_m=2$\,kHz, in the temperature range $T=13-20$\,K, where it is changing from a M to a V shape. In comparison, the temperature dependence of the V-type PRC for the $n$-type sample does not show a shape change (not shown here).

The full theory of SI is developed in the accompanying theoretical paper~\cite{TSI}. Here we briefly summarize the main results and apply the developed theory to the description of the experimental
results. The FE signal, $FE(f_m)$, is determined by the absolute value of the spin polarization at modulation frequency $f_m$. The spin dynamics for both types of carriers at small magnetic fields ($<1$\,T) is dominated by the hyperfine interaction with the host lattice nuclei. The resident carrier spin in each QD precesses in the random Overhauser field, while the nuclear spins are subject to the Knight field created by the carrier and experience the quadrupolar interaction induced by the strain in QDs~\cite{UrbaszekNuclReview13}. As a result, the spin polarization of an ensemble obeys a linear, but non-Markovian equation of motion. In the model of Merkulov, Efros, and Rosen~\cite{merkulov02} two-thirds of the electron spin polarization is lost during time scale of spin precession in the random Overhauser field, $\omega_N^{-1}$. The rest of spin polarization is destroyed at longer times, such as the nuclear spin correlation time $\tau_c$ related, e.g., to the nuclear quadrupole interaction, or the spin relaxation time $\tau_s$ unrelated to the hyperfine interaction.

The spin dynamics can be described by the Green's function $G(f_m)$ in the frequency domain. Theoretical analysis of the SI measurement protocol~\cite{TSI} shows that the SI signal is proportional to:
\begin{equation}
  FE(f_m)\propto Q G(f_m).
  \label{eq:L}
\end{equation}
Here the factor $Q$ is the probability of trion spin flip during its lifetime. According to the optical selection rules, trion excitation and recombination preserve the total spin along the excitation axis $z$. Therefore $Q$
determines the efficiency of spin polarization in QDs~\cite{Colton2012}. Importantly, it is sensitive to the external magnetic field~\cite{Fras13}, which ultimately gives rise to the different shape of PRC for different carriers. One can show, that~\cite{TSI}:
\begin{equation}
  Q=\frac{(\omega_N^\mathrm{T}\tau_0/\lambda^\mathrm{T})^2}{1+(\Omega_L^\mathrm{T}\tau_0)^2}+\frac{\tau_0}{\tau_s^\mathrm{T}},
\label{eq:Q}
\end{equation}
where $\omega_N^\mathrm{T}$ is the characteristic trion spin precession frequency in the fluctuations of the Overhauser field along the $z$ direction, $\lambda^\mathrm{T}$ is the parameter of the hyperfine interaction anisotropy of the trion, $\Omega_L^\mathrm{T}=g_{zz}^\mathrm{T}\mu_B B/\hbar$ is the trion spin precession frequency in the longitudinal external magnetic field with $g_{zz}^\mathrm{T}$ being the longitudinal $g$ factor and $\mu_B$ being the Bohr magneton, and $\tau_s^\mathrm{T}$ is the trion spin relaxation time unrelated to the hyperfine interaction. The laser repetition period $T_R$ is assumed to be much longer than the trion lifetime $\tau_0$.

The Green's function of the spin dynamics accounting for the finite nuclear spin correlation time is~\cite{Glazov_hopping}:
\begin{equation}
  G(f_m)=\frac{\tau_f\mathcal A}{1-\mathcal A\tau_f/\tau_c},
  \label{eq:Gzz}
\end{equation}
where $1/\tau_f=1/\tau_s+1/\tau_c-2\pi\i f_m$ and
\begin{equation}
  \mathcal A=\int d\bm \Omega_N \mathcal F(\bm \Omega_N)\frac{1+\Omega_{z}^2\tau_f^2}{1+\Omega^2\tau_f^2},
  \label{eq:A}
\end{equation}
with $\bm\Omega_N$ being the frequency of spin precession in the random Overhauser field in a single QD and $\bm\Omega=\bm\Omega_N+\bm\Omega_L$ being the total spin precession frequency in a single QD. The distribution of
the Overhauser field is Gaussian:
\begin{equation}
  \mathcal F(\bm \Omega_N)=\frac{\lambda^2}{\left(\sqrt{\pi}\omega_N\right)^3}\exp\left(-\frac{\Omega_{N,x}^2+\Omega_{N,y}^2}{\omega_N^2/\lambda^2}-\frac{\Omega_{N,z}^2}{\omega_N^2}\right),
\end{equation}
where $\omega_N$ is the typical spin precession frequency in the QD ground state and $\lambda$ describes the anisotropy of the hyperfine interaction. The latter is relevant ($\lambda>1$)~\footnote{$\lambda=1$ describes fully isotropic case.} for resident holes and for negatively charged trions, where the hole spin is unpaired, while the electrons form a singlet state with total spin zero, see inset in Fig.~\ref{fig:1}(a). We stress that the parameters of spin dynamics are different in the ground and excited states, so we used the superscript ``T'' in Eq.~\eqref{eq:Q} for the parameters of the trion spin dynamics.

\begin{table}[t]
\vspace{-0.3cm} \caption{Parameters of resident and photoexcited carriers spin dynamics, extracted from the fits in Fig.~\ref{fig:1}.}\label{tab:params}
\begin{ruledtabular}
\begin{tabular}{ccccccc}
& $g_{zz}$ & $\omega_N$, MHz & $\lambda$ & $\tau_c$, $\mu$s & $\tau_s$, $\mu$s & $\tau_s^\mathrm{T}$, $\mu$s \\
\hline
Electron & $-0.61$ & $70$ & $1$ & $0.2$ & $0.5$ & $<1$\\
Hole & $-0.45$ & $16$ & $5$ & $0.26$ & $5.2$ & $0.035$\\
\end{tabular}
\end{ruledtabular}
\end{table}

Equations~\eqref{eq:L}---\eqref{eq:A} allow us to describe the experimental data shown in Fig.~\ref{fig:1} and to determine the parameters of spin dynamics, which are summarized in
Tab.~\ref{tab:params}~\cite{params}. Below we describe qualitatively, how these parameters determine the PRC shape. As one can see from Eq.~\eqref{eq:L}, the frequency dependence of the FE is described by the Green's function $G(f_m)$. In general, this dependence is governed by the coupled spin dynamics of carrier and nuclei and differs from Eq.~\eqref{fit}. However in sufficiently strong magnetic fields, where $\Omega_L>\omega_{N}$ Eq.~\eqref{fit} is valid, provided $T_s=\tau_s$. In this case, the cut-off in the frequency dependence of the FE signal yields the spin relaxation time in the ground state, $\tau_s$. All other parameters of the spin dynamics can be extracted from the PRC at small frequency.

\begin{figure}[t]
\includegraphics[width=\columnwidth]{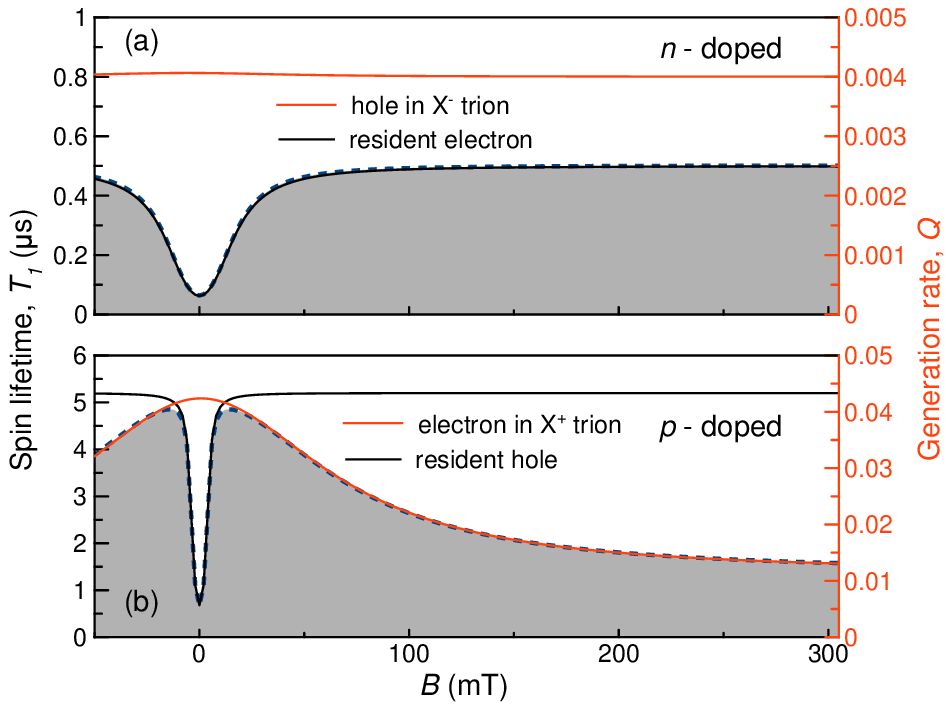}
\caption{Decomposition of the PRC (dashed blue lines, shaded for better visibility, arb. units) into magnetic field dependence of spin polarization of resident and photoexcited carriers:
(a) $n$-doped, (b) $p$-doped samples. Black curves give spin lifetime of resident carriers and red --- generation rate $Q$ determined by spin relaxation of the photoexcited carriers.
Parameters of the calculation are the same, as in Fig.~\ref{fig:1}.
}
\label{fig:3}
\end{figure}

At the smallest modulation frequencies $f_m\ll\tau_s^{-1}$, the FE signal does not depend on $f_m$ and the Green's function at zero modulation frequency determines the average spin relaxation time $T_1\equiv G(0)$. Therefore one
simply has $FE(f_m=0)\propto Q(B) T_1(B)$, where both factors depend on magnetic field. The decomposition of FE into these two contributions is shown in Fig.~\ref{fig:3}. Here the PRC curves calculated numerically (blue dashed
curves with gray filling) are shown together with the dependencies $T_1(B)$ (black lines) and $Q(B)$ (red lines) calculated for the same parameters as in Tab.~\ref{tab:params}.

The dip in the FE signals around zero magnetic field is determined by the dependence of $T_1$ on magnetic field. It is similar for both samples, as one can see in Fig.~\ref{fig:3}: the spin relaxation time increases with an increase of the longitudinal magnetic field, and then saturates. Its full width at half maximum is related to the longitudinal $g$ factor of the carrier and the characteristic nuclear field fluctuation, $\omega_N$, as $\sim\hbar\omega_N/(g_{zz}\mu_B)$~\cite{TSI}. Since the hyperfine interaction for electrons is stronger~\cite{Desfonds10}, the width of the $T_1$ curve is larger for the $n$-type sample, than for the $p$-type. We note that the saturation of the longitudinal spin relaxation time at large magnetic fields evidences the presence of spin relaxation mechanism unrelated to hyperfine interaction, which is described by the phenomenological time $\tau_s$. The deviations of the modelled behavior from the experimental one for the $n$-type sample in the intermediate field range may be caused by specific correlations in the nuclear dynamics, like quadrupole-induced spin relaxations, which are not included at this stage.

The ratio of the average spin relaxation time at large and zero magnetic fields is determined by the nuclear spin correlation time, $\tau_c$~\cite{TSI}:
\begin{equation}
  \frac{T_1(B=\infty)}{T_1(B=0)}=3+2\frac{\tau_s}{\tau_c}.
\end{equation}
Therefore the increase of $T_1$ in both samples by more than $3$ times indirectly evidences the finite hyperfine field correlation time. We note that the $\tau_c$ obtained from the fit is similar for both samples. This
indicates that the nuclear spin dynamics is related to the quadrupole interaction, which is similar for both samples~\cite{hackmann2015}.

Figure~\ref{fig:3} shows that the main difference between the FE dependence on $B$ for the two samples is the spin generation rate, or the trion spin-flip probability $Q(B)$. In positively charged QDs the trion consists of two heavy holes in the singlet state and an electron, see inset in Fig.~\ref{fig:1}(b). In this case, the trion spin relaxation at zero magnetic field is related to the hyperfine interaction of an unpaired electron spin with the host lattice nuclei in the QD. This is described by the first term in Eq.~\eqref{eq:Q}. With the increase of magnetic field, the electron spin gets effectively decoupled from the nuclear spin bath, and the trion spin-flip probability $Q$ decreases, as shown in Fig.~\ref{fig:3}(b). The width of the dependence $Q(B)$ is given by $2\hbar/(g_{zz}^\mathrm{T}\mu_B\tau_0)$. Using the $\tau_0=400$\,ps~\cite{GreilichPRB73}, we find the trion $g$ factor $g_{zz}^\mathrm{T}=-0.4$, which is close to the bare electron $g$ factor. Moreover, assuming that the characteristic nuclear field is the same as for $n$-type QDs $\omega_N^\mathrm{T}=70$\,MHz, we find the trion spin relaxation time unrelated to the hyperfine interaction $\tau_s^\mathrm{T}=35$\,ns.

For negatively charged QDs the trion consists of two electrons in the singlet state and a hole with unpaired spin, as shown in the inset in Fig.~\ref{fig:1}(a). In this case, the spin dynamics of trion is determined by hole spin relaxation. The hyperfine interaction of negatively charged trion is weaker than of positively charged trion. As a result the trion spin relaxation in $n$-type QDs is unrelated to hyperfine interaction and $Q$ does not depend on $B$, see the red curve in Fig.~\ref{fig:3}(a). Mathematically the second term in Eq.~\eqref{eq:Q} dominates in this case, which allows us to estimate $\tau_s^\mathrm{T}<1~\mu$s in the $n$-type sample.

Interestingly, the dependence $Q(B)$ can change with a temperature increase, provided $\tau_s^\mathrm{T}$ depends on $T$. This takes place, for example, for the phonon-assisted spin relaxation mechanism, although the precise origin of $\tau_s^\mathrm{T}$ for both types of quantum dots is not fully clear so far. The spin relaxation speeds up with an increase of temperature, so the dependence $Q(B)$ can become flat even for $p$-type QDs. This explains the change of the shape of the FE signal from M-like to V-like, shown in Fig.~\ref{fig:2}.

It is instructive to compare the SI technique with the spin noise technique~\cite{VSZRev,Oestreich-review}. Indeed, both give access to the Fourier transform of the spin dynamics Green's function, which can be measured for different magnetic fields. In fact, one can show, that the spin noise spectrum is proportional to $\Re G_{zz}(f_m)$~\cite{TSI}. Thus, it is characterized by the same parameters as given in Tab.~\ref{tab:params}, but in a different way, which makes these two approaches complementary.

To conclude, we show that the SI measurement gives access to various parameters of the spin dynamics not only of resident charge carriers but also of the photoexcited electron-hole complexes.

\begin{acknowledgments}
We acknowledge the financial support by the Deutsche Forschungsgemeinschaft in the frame of the International Collaborative Research Center TRR 160 (Project A5), partial support by the Russian Foundation for Basic Research
(Grant No. 15-52-12012), and Basis Foundation.
\end{acknowledgments}

\bibliographystyle{apsrev}

\end{document}